\documentclass[twoside]{dis09}
\usepackage[latin1]{inputenc}
\usepackage[dvips]{graphicx,epsfig,color}
\usepackage{amssymb,amsmath,array}

\begin{document}
\title{Photon + jets at D\O }

%***********************************************************************
% AUTHORS INFORMATION AREA
%***********************************************************************
%\author{Lars Sonnenschein$^1$
\author{Lars Sonnenschein
%
% Optional short acknowledgment: remove next line if non-needed
%\thanks{This is an optional funding source acknowledgment.}
%
% DO NOT MODIFY THE FOLLOWING '\vspace' ARGUMENT
\vspace{.3cm}\\
%
% Addresses and institutions (remove "1- " in case of a single institution)
%1- RWTH Aachen - III. institute A \\
RWTH Aachen - Institute III A \\
Physikzentrum 52056 Aachen - Germany
%
% Remove the next three lines in case of a single institution
%\vspace{.1cm}\\
%2- School of Second Author - Dept of Second Author \\
%Address of Second Author's school - Country of Second Author's school\\
}
%***********************************************************************
% END OF AUTHORS INFORMATION AREA
%***********************************************************************

\maketitle

\begin{abstract}
Photon plus jet production has been studied by the D\O\ experiment 
in Run~II of the Fermilab Tevatron Collider at a centre of mass energy of $\sqrt{s}=1.96$~TeV.
Measurements of the inclusive photon, inclusive photon plus jet, photon plus heavy flavour 
jet cross sections and double parton interactions in photon plus three jet events
are presented. They are based on integrated luminosities between 0.4~fb$^{-1}$ and 1.0~fb$^{-1}$.
The results are compared to perturbative QCD calculations in various approximations.
\end{abstract}

\section{Introduction}

Photons originating from the hard subprocess and produced during fragmentation 
contribute to photon cross sections in hadron-hadron collisions.  
The contribution of fragmentation photons can 
be significantly reduced by isolation requirements. 
Thus, isolated photon cross sections are sensitive to
the dynamics of the hard subprocess, to the strong coupling constant $\alpha_s$ and to the parton
distribution functions (PDF's) of the colliding hadrons. In the following isolated photon and 
photon plus jet cross section measurements from the D\O\ experiment are presented~\cite{d0pj}
as well as a measurement of double parton interactions in photon plus three jet events.

\section{Inclusive photon cross section}

D\O\ has measured the inclusive photon cross section~\cite{ip} based on an integrated 
luminosity of 380~pb$^{-1}$.
Photon candidates are defined as clusters of electromagnetic (EM) calorimeter cells within a cone
of radius $R=0.2$ in the space of pseudorapidity $\eta$ and azimuthal angle $\phi$,
if more than 95\% of the detected energy is located in the EM layers of the calorimeter
%
%\begin{wrapfigure}{h}{1.0\columnwidth}
\begin{figure}[h]
\vspace*{-3ex}
\centerline{
\includegraphics[width=0.47\columnwidth]{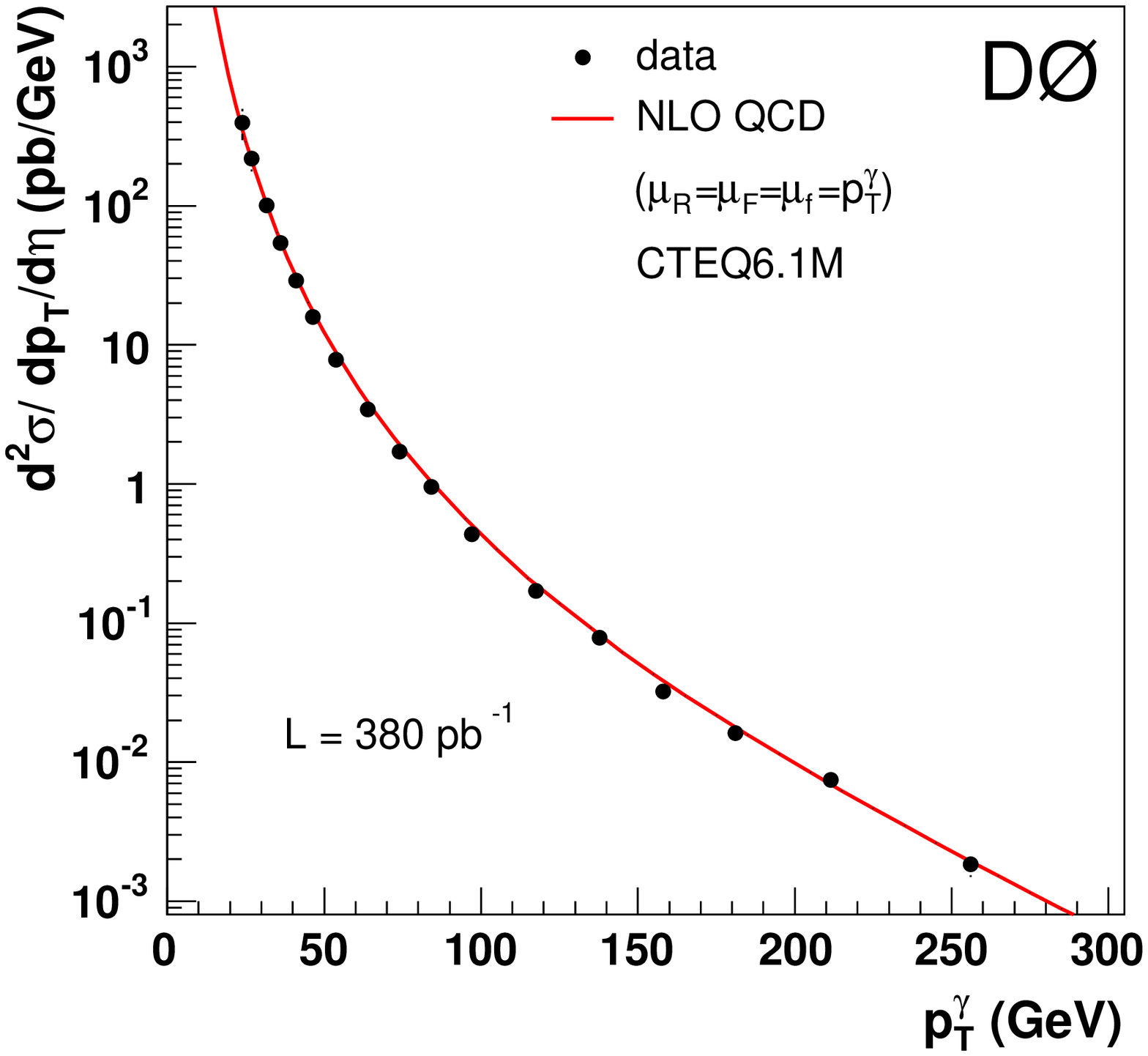}
\includegraphics[width=0.5\columnwidth]{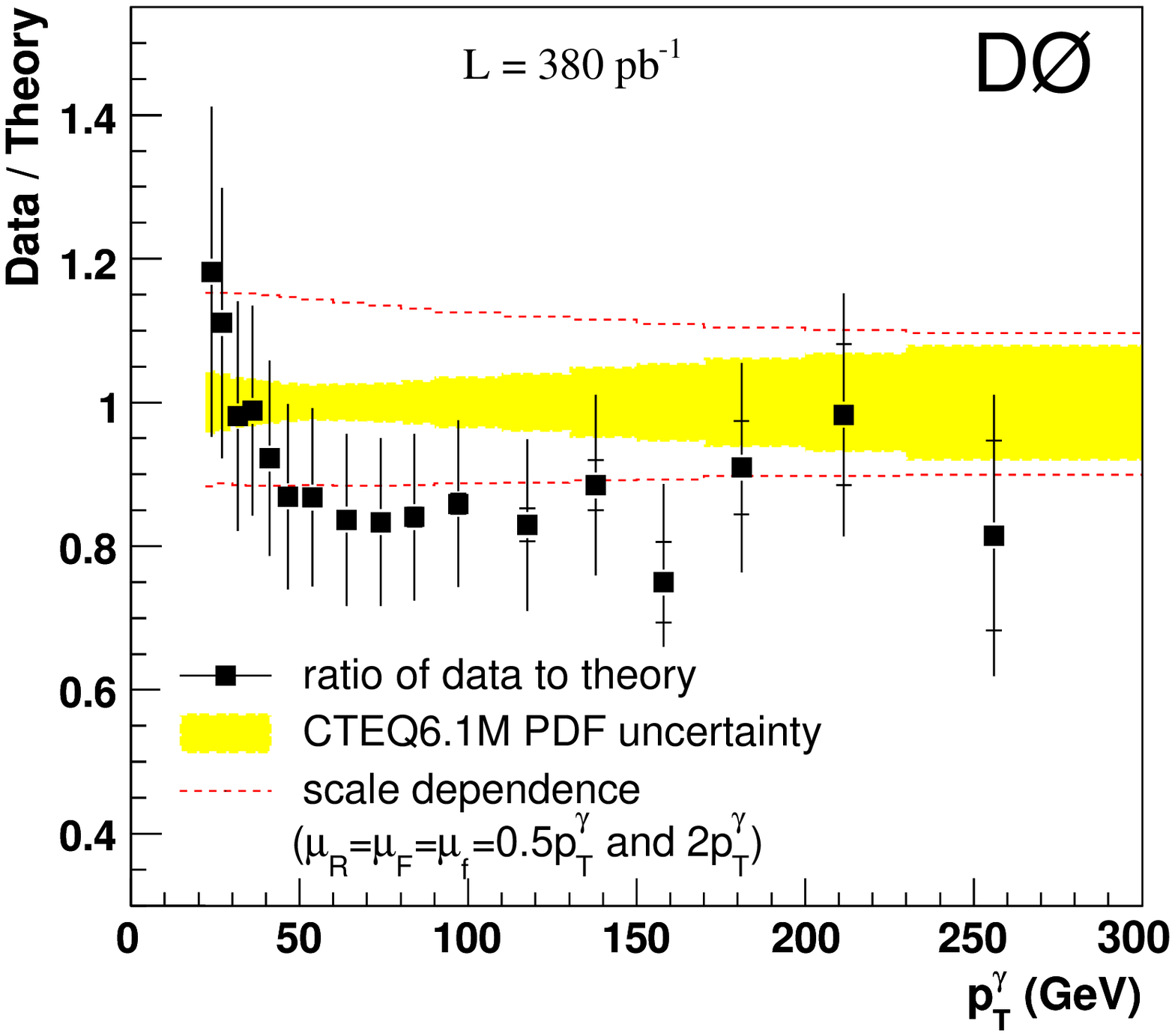}}
\vspace*{-3ex}
\caption{Double differential inclusive photon cross section (left) and data over theory 
ratio (right) in comparison to the prediction of JetPhox.}\label{Fig:ip}
%\end{wrapfigure}
\vspace*{-1ex}
\end{figure}
\pagebreak
and the probability for a track match is below 0.1\%.
As isolation criterion the transverse energy not associated to the photon in a cone of radius 
$R=0.4$ around the photon direction has to be less than 0.10 times the energy of the photon.
Backgrounds from cosmics and electrons from $W$ boson decays are vetoed by a missing transverse
energy requirement of $E_T\!\!\!\!\!\!\!/\;\, < 0.7p_T^{\gamma}$.
Electromagnetic cluster and track information is fed into a neural network (NN) to
increase the photon purity further. 
Central photons ($|\eta|<0.9$) with a transverse momentum above 23~GeV are selected.
The differential cross section is compared to the NLO prediction of 
JetPhox~\cite{jpx1}\cite{jpx2}\cite{px} in Fig. \ref{Fig:ip}.
The results are still consistent with theory but a shape similar to previous observations
of UA2 and CDF emerges.

\section{Photon plus jet cross section}

The inclusive photon plus jet cross section~\cite{pj} has been measured by D\O\
based on an integrated luminosity of 1.0~fb$^{-1}$.
%Photon candidates are defined as clusters of electromagnetic (EM) calorimeter cells within a cone
%of radius $R=0.2$ in the space of pseudorapidity $\eta$ and azimuthal angle $\phi$,
Photon candidates have to deposit at least 96\% of the detected energy in the EM layers of the 
calorimeter. 
%and the probability for a track match is below 0.001.
As isolation criterion the transverse energy not associated to the photon in a cone of radius 
$R=0.4$ around the photon direction has to be less than 0.07 times the energy of the photon.
The missing transverse energy requirement is 
$E_T\!\!\!\!\!\!\!/\;\, < 12.5~\mbox{GeV} + 0.36p_T^{\gamma}$.
Electromagnetic cluster and track information is fed into a neural network to
increase the photon purity further. 
Central photons ($|\eta|<1.0$) with a transverse momentum above 30~GeV are selected.
Jets are defined in the energy scheme by the Run~II midpoint cone 
algorithm~\cite{r2} with a radius of $R=0.7$.
The jets with a transverse momentum above 15~GeV are selected in the central 
($|\eta^{\mbox{\scriptsize jet}}|<0.8$) or the forward ($1.5<|\eta^{\mbox{\scriptsize jet}}|<2.5$) 
region.
Finally, the photon and the leading hadronic jet have to be separated by 
$\Delta R(\gamma,\mbox{\scriptsize jet})>0.7$.

The triple differential cross section is measured as a function of transverse photon momentum and 
jet and photon rapidities in four bins, distinguished by central and forward jets on the one 
hand and same/opposite side jet photon rapidities on the other hand. These four kinematic intervals
probe different momentum phase space regions of the two initial interacting partons.
%
%
%\begin{wrapfigure}{r}{1.0\columnwidth}
\begin{figure}[b]
%\centerline{
\unitlength 1cm
\begin{picture}(9.0,5.0)
\put(0.5,0.3){\includegraphics*[width=0.38\columnwidth, bb=283 271 535 517, clip=true]{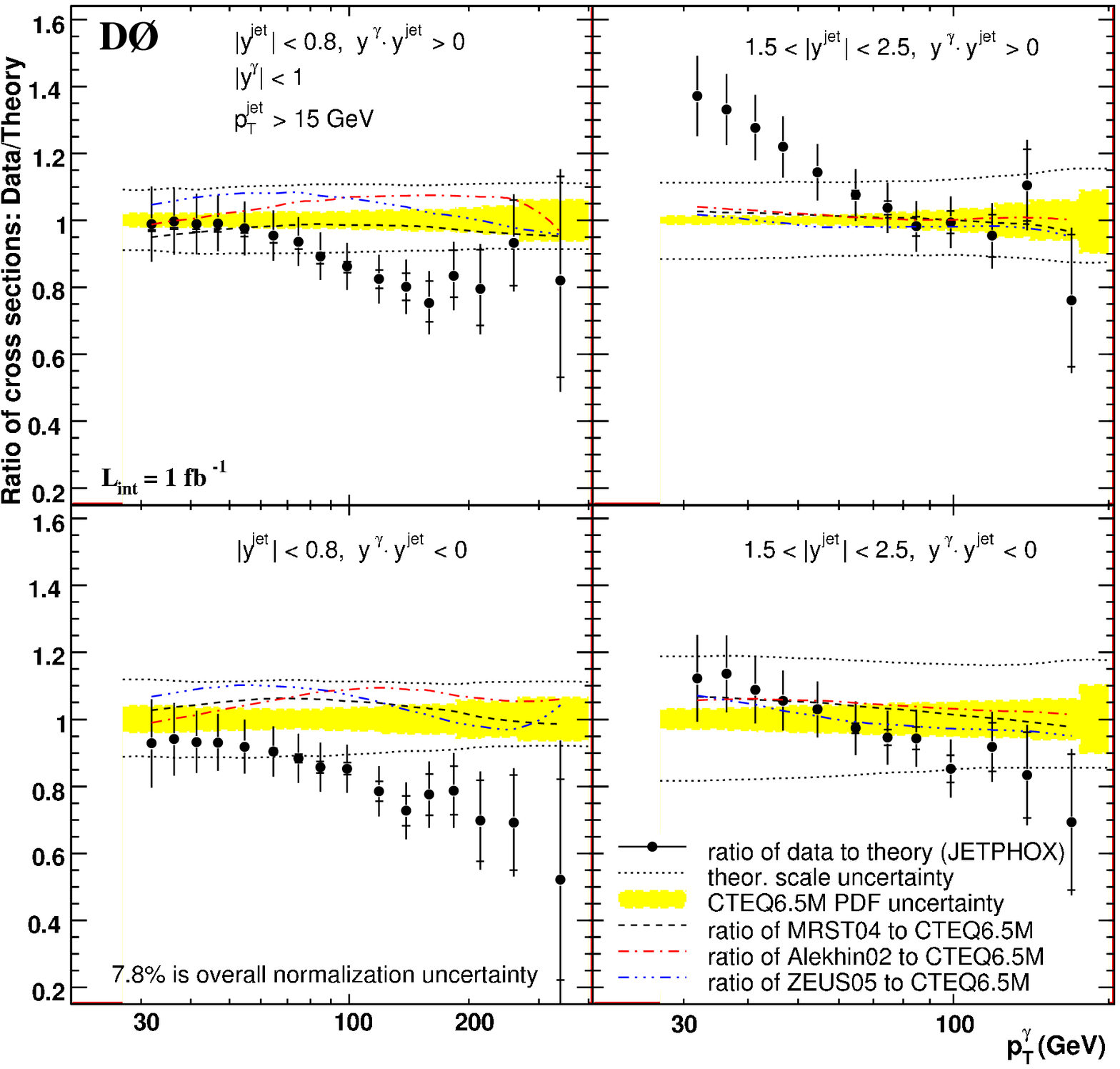}}
\put(0.7,0.6){\includegraphics*[width=0.33\columnwidth, bb=290 38 510 115, clip=true]{Sonnenschein_Lars.fig3.eps}}
\put(0.7,4.7){\includegraphics*[width=0.05\columnwidth, bb=45 478 78 505, clip=true]{Sonnenschein_Lars.fig3.eps}}
\put(-0.25,0.3){\includegraphics*[width=0.049\columnwidth, bb=0 271 32 525, clip=true]{Sonnenschein_Lars.fig3.eps}}
\put(0.5,-0.4){\includegraphics*[width=0.385\columnwidth, bb=283 0 535 30, clip=true]{Sonnenschein_Lars.fig3.eps}}
\put(6.65,0.3){\includegraphics[width=0.565\columnwidth, bb=283 187 550 356, clip]{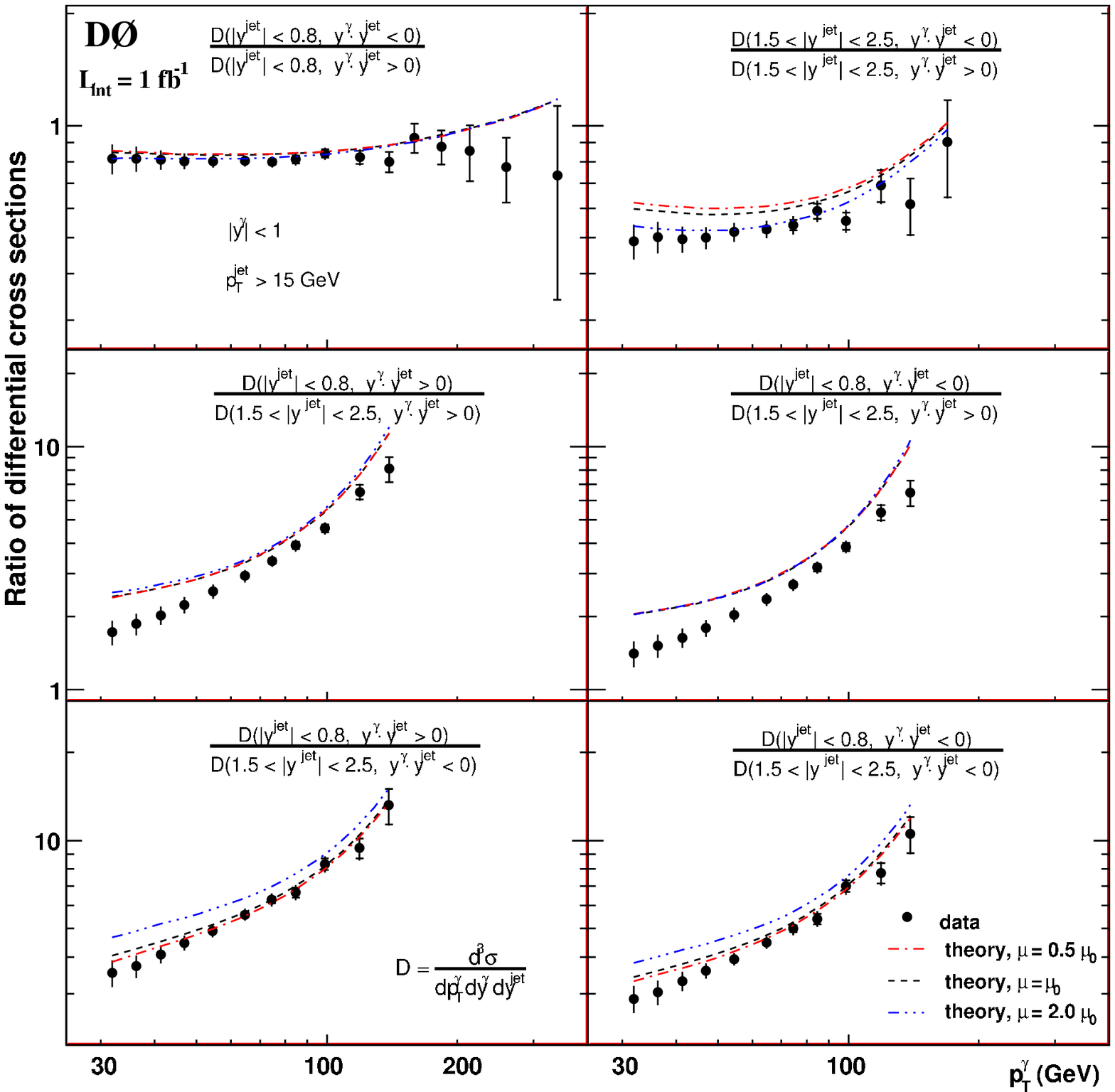}}
\put(10.8,0.6){\includegraphics[width=0.2\columnwidth, bb=425 30 528 90, clip]{Sonnenschein_Lars.fig4.eps}}
\put(6.92,3.9){\includegraphics[width=0.1\columnwidth, bb=40 470 97 516, clip]{Sonnenschein_Lars.fig4.eps}}
\put(6.23,0.37){\includegraphics[width=0.025\columnwidth, bb=20 187 32 356, clip]{Sonnenschein_Lars.fig4.eps}}
\put(6.55,-0.275){\includegraphics[width=0.57\columnwidth, bb=280 0 550 17, clip]{Sonnenschein_Lars.fig4.eps}}
\end{picture}
%}
\caption{Data over theory ratio and ratio of two photon and jet rapidity bins.}\label{Fig:pj}
%\end{wrapfigure}
\end{figure}
\pagebreak
Fig. \ref{Fig:pj} (left) 
shows the data versus theory (JetPhox) ratio as a function of photon transverse momentum 
for forward jets and same side photon and jet events. 
The theory is not able to describe the shape of the data over the whole
measured range of photon rapidities. The shape is similar to previous observations of UA2 and CDF.
The right plot shows the ratio of the cross section of central
jets with opposite side photon rapidities over the cross section of forward jets with same side 
photon rapidities for both, data and theory. The ratio has the advantage of reduced uncertainties
due to correlated errors. The cross section ratio reveals quantitative 
disagreement between data and theory.

\section{Inclusive photon plus heavy flavour jet cross section}
\begin{figure}[h]
\unitlength 1cm
\begin{picture}(9.0,7.15)
\put(-0.65,-0.7){\includegraphics[width=0.62\columnwidth]{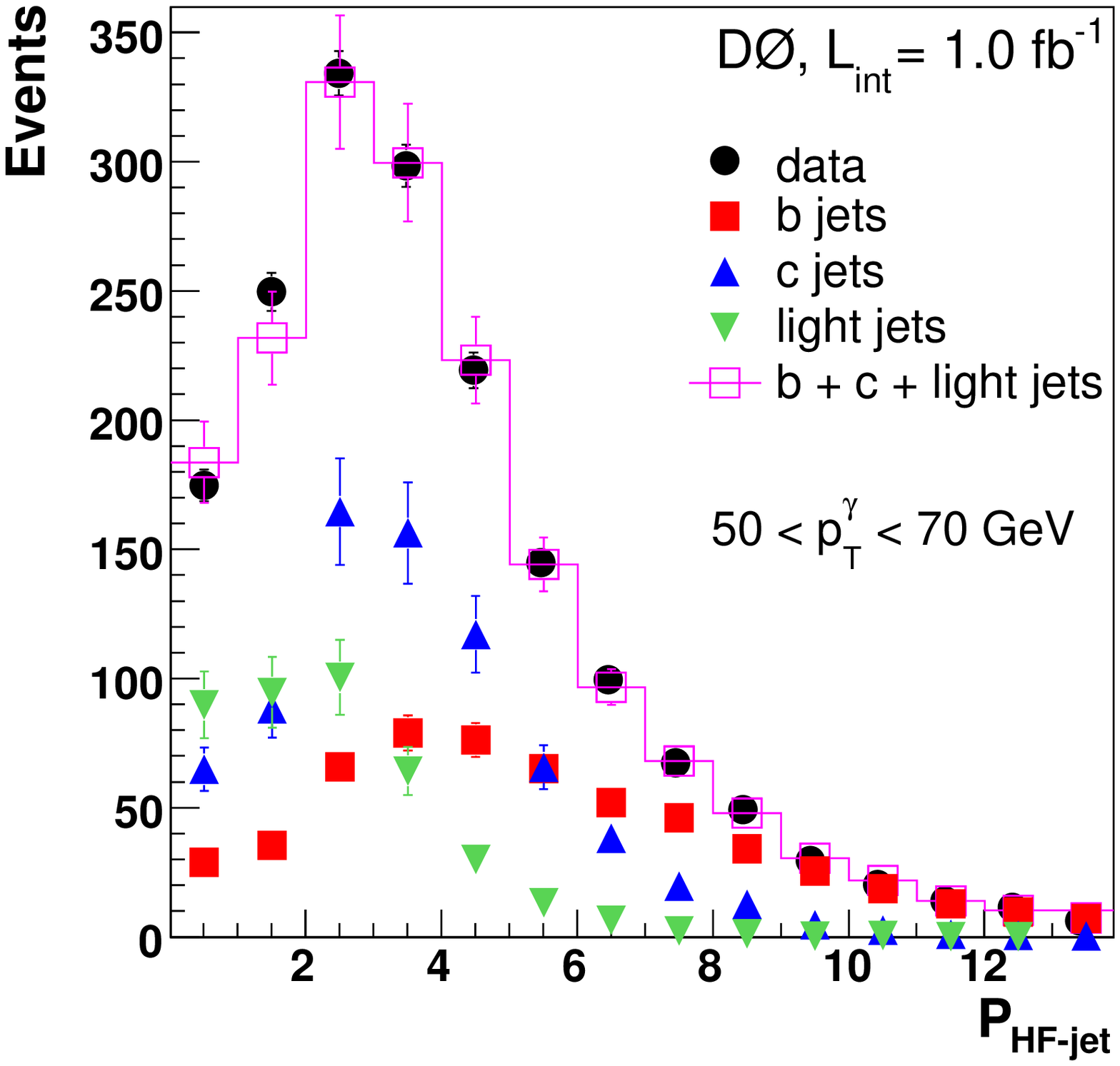}}
\put(7.25,0.0){\includegraphics[width=0.5\columnwidth]{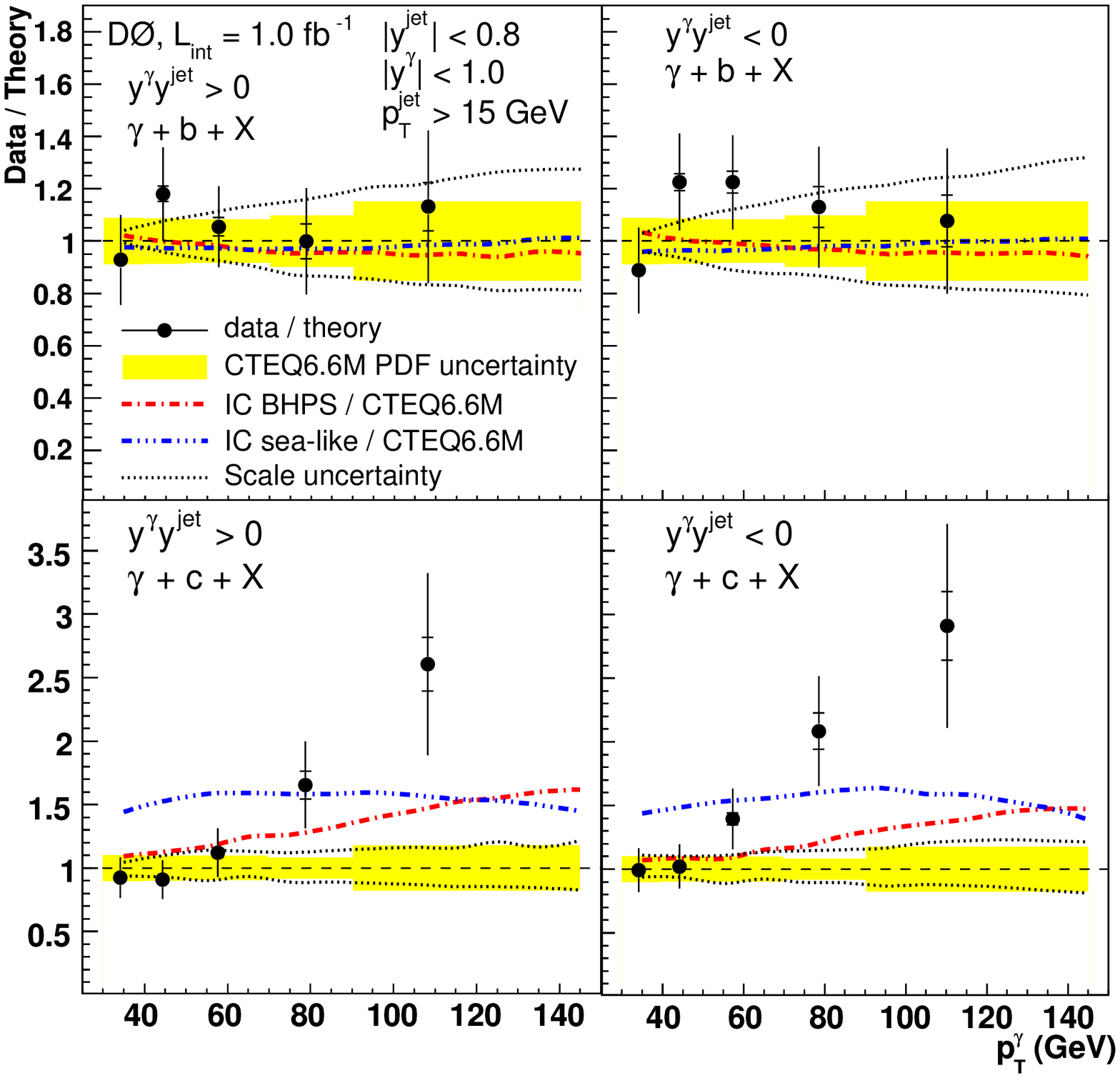}}
\end{picture}	
\caption{Jet flavour templates and data over theory ratio in comparison to theory.}\label{Fig:phj}
%\end{wrapfigure}
\end{figure}

D\O\ has  measured the inclusive photon plus heavy flavour jet cross section~\cite{phj} 
making use of an integrated luminosity of 1.0~fb$^{-1}$.
One isolated photon with a transverse momentum above 30~GeV has to be found in the rapidity
range of $|y_{\gamma}|<1.0$. 
Electromagnetic cluster and track information is fed into a neural network to
improve the photon purity.
Backgrounds from cosmics and electrons from $W$ boson decays are vetoed by a missing transverse
energy requirement of $E_T\!\!\!\!\!\!\!/\;\, < 0.7p_T^{\gamma}$.
D\O\ Run~II jets~\cite{r2} with a cone radius of 0.5 and 
a transverse momentum above 15~GeV are considered
in the central pseudorapidity range of $|y_{\mbox{\scriptsize jet}}|<0.8$.
The leading momentum jet has to have at least two tracks and a neural network 
which exploits the longer lifetimes of heavy flavoured hadrons is applied
to enhance the heavy flavour jet content of the considered events. 
Same side and opposite side photon and jet rapidity events are treated separately.
The fractional contributions of $b$ and $c$ jets are determined by fitting templates
(Fig. \ref{Fig:phj} left) of a function 
$P_{\mbox{\scriptsize HF-jet}} = -\ln\prod_i P^i_{\mbox{\scriptsize track}}$ 
to the data, where $P^i_{\mbox{\scriptsize track}}$ is the probability that a track originates
from the primary vertex. Jets from $b$ quarks have typically large values of 
$P_{\mbox{\scriptsize HF-jet}}$. Fig. \ref{Fig:phj} (right) shows the data over theory
%
%\begin{wrapfigure}{r}{1.0\columnwidth}
\pagebreak
ratio as a function of the photon transverse momentum 
of the $b$ quark plus photon production (top) and $c$ quark plus photon production 
cross sections in the same side (left) and opposite side (right) photon and jet rapidity bins.
The next-to-leading order predictions~\cite{hfth} are based on techniques to calculate the
cross section analytically~\cite{hfth2}.
While the prediction agrees with the measured cross section for $b$ quark plus photon production
over the whole range of photon transverse momenta, the prediction underestimates the measured
cross section for $c$ quark plus photon production for photon transverse momenta above 70~GeV.

\section{Double parton interactions in photon plus three jet events}
The double parton scattering in photon plus three jet events has been measured 
by D\O~\cite{dps} making use of an integrated luminosity of 1.0~fb$^{-1}$.
This measurement provides complementary information about the proton structure, namely
the spatial distribution of partons inside the proton. 
Possible parton-parton correlations and an impact
on the proton parton distribution functions (PDF's) can be investigated.
Events with an isolated photon in the transverse momentum range between 60 and 80~GeV,
a leading jet with a transverse momentum above 25~GeV and two further jets with a
transverse momentum above 15~GeV are selected.
The main background arises from single parton scattering events with additional jets from 
initial and final state radiation. The cross section for double parton scattering can be expressed
as $\sigma_{\mbox{\scriptsize DP}}=m\sigma_A\frac{\sigma_B}{2\sigma_{\mbox{\scriptsize eff}}}$,
where $\sigma_{A,B}$ are the cross sections of the process A and B.
$\sigma_{\mbox{\scriptsize eff}}$ characterises the size of the effective interaction region
and the term $\sigma_B/2\sigma_{\mbox{\scriptsize eff}}$ gives the probability of a second 
interaction $B$, given that a first interaction $A$ has already taken place.
In this analysis the permutation factor $m$ equals to two since the processes $A$ and $B$ 
can be distinguished due to the photon.
The measurement is done in three transverse momentum bins of the second jet from 15 to 20,
20 to 25 and 25 to 30~GeV using data driven techniques based on the different transverse momentum 
spectra between jets of dijet events and those of single parton scattering events with jet radiation.
%
%\begin{wrapfigure}{r}{1.0\columnwidth}
\begin{figure}[h]
\vspace*{-3ex}
\centerline{
\includegraphics[width=0.5\columnwidth]{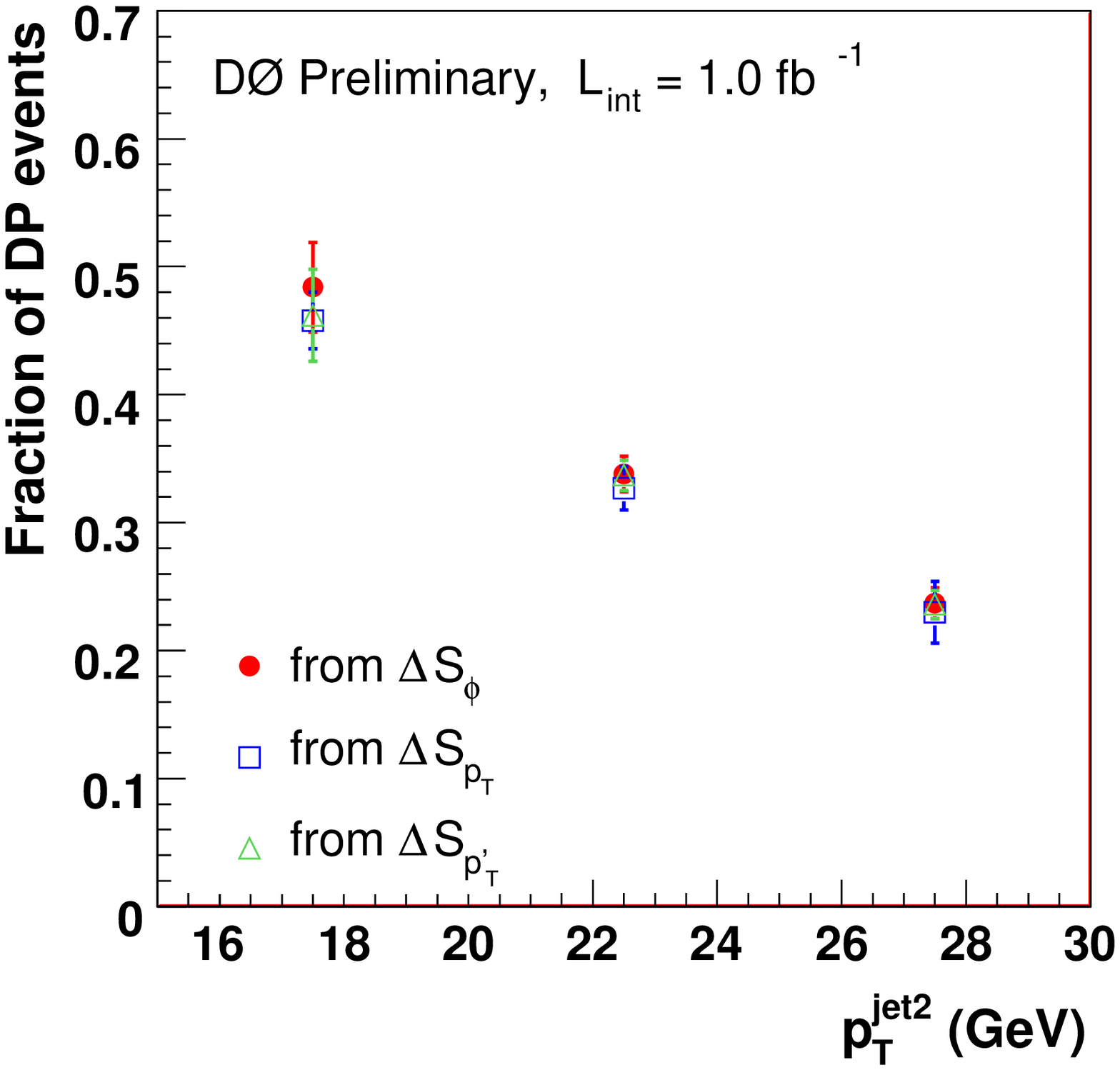}
\includegraphics[width=0.5\columnwidth]{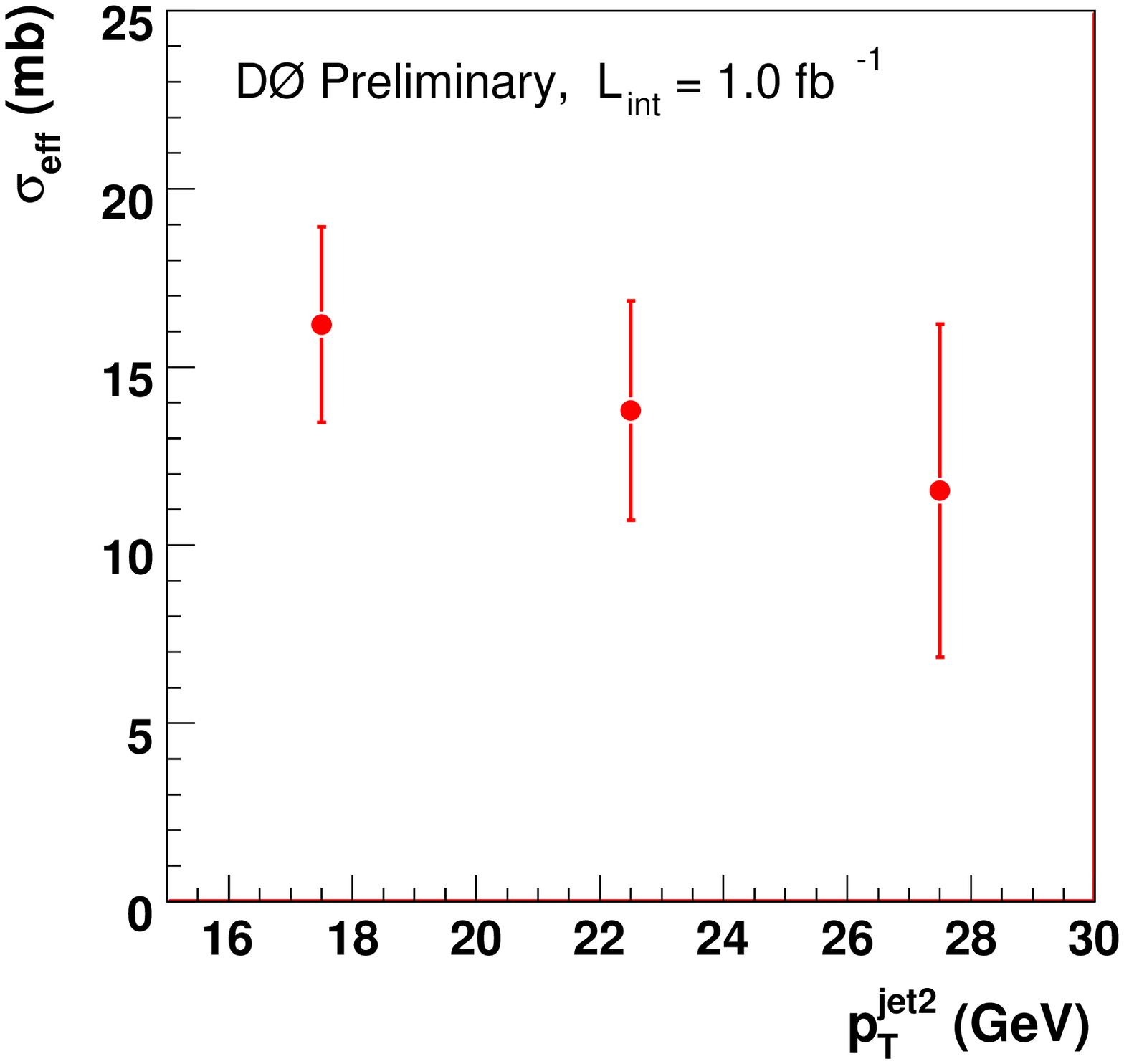}}
\vspace*{-2ex}
\caption{Fraction of double parton events (left) and $\sigma_{\mbox{\scriptsize eff}}$
(right) as a function of the second transverse momentum jet.}\label{Fig:dps}
%\end{wrapfigure}
\end{figure}
\pagebreak
The discriminating variables 
$S_{\phi}=\frac{1}{\sqrt{2}}\sqrt{\left( \frac{\Delta\phi(\gamma,i)}{\delta\phi(\gamma,i)} \right)^2 + \left( \frac{\Delta\phi(j,k)}{\delta\phi(j,k)}\right)^2}$, 
$S_{p_T}=\frac{1}{\sqrt{2}}\sqrt{\left( \frac{|\vec{P_T}(\gamma,i)|}{\delta P_T(\gamma,i)} \right)^2 + \left( \frac{|\vec{P_T}(j,k)|}{\delta P_T(j,k)}\right)^2}$ and 
$S_{p'_T}=\frac{1}{\sqrt{2}}\sqrt{\left( \frac{|\vec{P_T}(\gamma,i)|}{|\vec{P}_T^{\gamma}|+|\vec{P}_T^i|} \right)^2 + \left( \frac{|\vec{P_T}(j,k)|}{|\vec{P}_T^j|+|\vec{P}_T^k|}\right)^2}$ 
are constructed, where each object can be either the photon 
or one of the three jets. Essentially the variables correspond to the quadratic sum
of the object pair azimuthal angle difference significances, the object pair 
transverse momentum significances and the object pair transverse momentum vectors over scalar sum, 
respectively. Finally the observable
$\Delta S = \Delta\phi(p_T^{\gamma,\; \mbox{\scriptsize jet$_i$}},
        p_T^{\mbox{\scriptsize jet$_j$, jet$_k$}})$ is computed for the object pair
combination which minimises $S$. 
The distributions of the distinguishing $\Delta S$ variables are compared to two sets
of photon plus three jet data with one vertex events which differ in signal fractions.
As it turned out single parton scattering events are mostly ($\gtrsim 90\%$) concentrated in
the region $\Delta S>2.0$.
Fig. \ref{Fig:dps} (left) shows the fraction of measured double parton scattering events
which is decreasing with increasing second jet transverse momentum. The right plot shows
the measured effective cross section $\sigma_{\mbox{\scriptsize eff}}$. The measurements in the 
different second jet transverse momentum bins are consistent with each other within errors and an
average effective cross section $<\sigma_{\mbox{\scriptsize eff}}>=15.1\pm1.9$~mb is obtained.
This value is consistent with previous measurements of UA2 and CDF.

%%%%%%%%%%%%%%%%%%%%%%%%%%%%%%%%%%%%%%%%%%%%%%%%%%%%%%%%%%%%%%%%%%%%%%%%%%%%%%%%%%%%

\section{Acknowledgments}

Many thanks to the staff members at Fermilab and collaborating institutions. 
This work has been supported by the DOE and NSF (USA); CEA and CNRS/IN2P3 (France); 
FASI, Rosatom and RFBR (Russia);
CNPq, FAPERJ, FAPESP and FUNDUNESP (Brazil); DAE and DST (India); Colciencias (Columbia); 
CONACyT (Mexico); KRF and KOSEF (Korea); CONICET and UBACyT (Argentina); FOM (The Netherlands);
STFC (United Kingdom); MSMT and GACR (Czech Republic); CRC Program, CDF, NSERC and WestGrid Project 
(Canada); BMBF, DFG and the Alexander von Humboldt Foundation (Germany); SFI (Ireland); 
The Swedish Research Council (Sweden); and CAS and CNSF (China).

\section{Bibliography}

% ****************************************************************************
% BIBLIOGRAPHY AREA
% ****************************************************************************

\begin{footnotesize}
% IF YOU DO NOT USE BIBTEX, USE THE FOLLOWING SAMPLE SCHEME FOR THE REFERENCES
% ----------------------------------------------------------------------------

% ----------------------------------------------------------------------------

% IF YOU USE BIBTEX,
% - DELETE THE TEXT BETWEEN THE TWO ABOVE DASHED LINES
% - UNCOMMENT THE NEXT TWO LINES AND REPLACE 'Name_Of_Your_BibFile'

%\bibliographystyle{unsrt}
%\bibliography{Name_Of_Your_BibFile}
% example of Name_Of_Your_BibFile.bib
% @Article{Turcato:2006ch,
%      author    = "Turcato, M.",
%  collaboration = "ZEUS and H1",
%      title     = "Lepton flavour violation and charmonium physics at HERA",
%      journal   = "Nucl. Phys. Proc. Suppl.",
%      volume    = "162",
%      year      = "2006", 
%      pages     = "283-287",
%      SLACcitation  = "%%CITATION = NUPHZ,162,283;%%"
% }
% 
% @Unpublished{Gogitidze:2007du,
%      author    = "Gogitidze, N.",
%  collaboration = "H1", 
%      title     = "Prompt photons and particle momentum distributions at
%                   HERA", 
%      year      = "2007",
%      note    = "hep-ex/0701033",
%      SLACcitation  = "%%CITATION = HEP-EX 0701033;%%"
% }

\end{footnotesize}

% ****************************************************************************
% END OF BIBLIOGRAPHY AREA
% ****************************************************************************

\end{document}